\documentstyle[psfig]{mn}

\title[The X-ray Spectrum and Light Curve of SN~1995N]
  {The X-ray Spectrum and Light Curve of Supernova 1995N}

\author[D. W. Fox et al.]{
            D.~W.~Fox,$^1$ W.~H.~G.~Lewin,$^1$ 
            A.~Fabian,$^2$ K.~Iwasawa,$^2$ R.~Terlevich,$^3$
 \newauthor H.~U.~Zimmermann,$^4$ B.~Aschenbach,$^4$
            K.~Weiler,$^5$ S.~Van~Dyk,$^6$
            R.~Chevalier,$^7$ 
 \newauthor R.~Rutledge,$^8$ H.~Inoue,$^9$ S.~Uno$^9$\\
 $^1$MIT Center for Space Research, 77 Massachusetts Ave \#37-627,
     Cambridge, MA 02139-4307, USA\\
 $^2$Institute of Astronomy, Madingley Road, Cambridge CB3 0HA \\
 $^3$Royal Greenwich Observatory, Madingley Road, Cambridge CB3 0EZ \\
 $^4$Max-Planck-Institut f\"ur Extraterrestrische Physik,
     Postfach 1312, D-85741 G\"arching, Germany \\
 $^5$Naval Research Laboratory, Washington, DC 20375-5300, USA \\
 $^6$IPAC/Caltech 100--22, Pasadena, CA 91115, USA \\
 $^7$Department of Astronomy, University of Virginia, Charlottesville,
            VA 22903-0818, USA \\
 $^8$California Institute of Technology MC 220--47, Pasadena, CA 91125, USA \\
 $^9$Institute of Space and Astronautical Science, Sagamihara, Japan
}

\date{Accepted date. Received date.}
\pagerange{\pageref{firstpage}--\pageref{lastpage}}
\pubyear{2000}

\newcommand{\rosat}{\textit{ROSAT}}
\newcommand{\rosatz}{\textit{ROSAT\/}}
\newcommand{\asca}{\textit{ASCA}}
\newcommand{\ascaz}{\textit{ASCA\/}}

\newcommand{\chandra}{\textit{Chandra}}
\newcommand{\xmm}{\textit{XMM-Newton}}
\newcommand{\etal}{et al.}
\newcommand{\snn}{\mbox{SN~1995N}}
\newcommand{\mcg}{\mbox{MCG~$-$02-38-017}}
\newcommand{\nhcol}{\mbox{$N_{\rm H}$}}
\newcommand{\pcmsq}{\mbox{cm$^{-2}$}}
\newcommand{\ergcms}{ergs cm$^{-2}$ s$^{-1}$}
\newcommand{\ergsec}{ergs s$^{-1}$}
\newcommand{\kmsec}{km s$^{-1}$}
\newcommand{\Lx}{\mbox{$L_{\rm X}$}}
\newcommand{\rah}{\mbox{$^{\rm h}$}}
\newcommand{\ram}{\mbox{$^{\rm m}$}}
\newcommand{\ras}{\mbox{$^{\rm s}\!$.}}
\newcommand{\dcd}{\mbox{$^{\circ}$}}
\newcommand{\dcm}{\mbox{'}}
\newcommand{\dcs}{\mbox{''$\!$.}}
\newcommand{\dgsp}{\mbox{~$\,$}}

\newcommand{\multc}[2]{\multicolumn{#1}{c}{#2}}

\newcommand{\ccol}[1]{\multicolumn{1}{c}{#1}}

\begin{document}
\label{firstpage}
\maketitle

\begin{abstract}
We report on multi-epoch X-ray observations of the Type~IIn (narrow
emission line) supernova \snn\ with the \rosatz\ and \ascaz\
satellites.  The January~1998 \ascaz\ X-ray spectrum is well fitted by
a thermal bremsstrahlung ($kT \sim 10$~keV, $\nhcol \sim 6\times
10^{20}$~\pcmsq) or power-law ($\alpha \sim 1.7$, $\nhcol \sim
10^{21}$~\pcmsq) model.  The X-ray light curve shows evidence for
significant flux evolution between August~1996 and January~1998: the
count rate from the source decreased by 30\% between our August~1996
and August~1997 \rosatz\ observations, and the X-ray luminosity most
likely increased by a factor of $\sim$2 between our August~1997
\rosatz\ and January~1998 \ascaz\ observations, although evolution of
the spectral shape over this interval is not ruled out.  The high
X-ray luminosity, $\Lx \sim 10^{41}$ \ergsec, places \snn\ in a small
group of Type~IIn supernovae with strong circumstellar interaction,
and the evolving X-ray luminosity suggests that the circumstellar
medium is distributed inhomogeneously.

\end{abstract} 


\section{Introduction}
\label{sec:intro}
A small number of supernovae (SNe) have been detected in X-rays in the
near aftermath ($\sim$years) of their explosions.  SN~1978K, SN~1980K,
SN~1986J, SN~1987A, and SN~1993J were found before 1995 (see Schlegel
1995 for a review); more recent detections include SN~1979C, SN~1988Z,
and SN~1994I (Immler, Pietsch, \& Aschenbach 1998a,b) and SN~1999em
(Fox \& Lewin 1999).  Here we report on observations of \snn\ (Pollas
1995).

Supernova X-ray emission depends on the density structure of the
progenitor's stellar wind as well as the structure of the SN ejecta,
through which the initial shock passes.  After the shock wave emerges
from the star, the characteristic velocity is $\sim$10$^4$ \kmsec\ and
the density distribution in the outer parts of the star can be
approximated by a power-law in radius, $\rho \propto r^{-n}$, where
the value of $n$ ranges from $\sim$7 to $\sim$20 (models for SN1987A
have $n \sim 10$; Matzner \& McKee 1999).  The shock then propagates
into the circumstellar material, typically a slow-moving wind with a
density that decreases with the inverse square of the radius, $\rho =
\dot{M} / 4 \pi r^2 v_w$, where $\dot{M}$ is the stellar wind
mass-loss rate and $v_w$ is the wind velocity (typically $\sim$10
\kmsec).  The collision between the stellar ejecta and the
circumstellar material produces a second `reverse' shock at $\sim$1000
\kmsec in the stellar ejecta (which is expanding at $\sim$10$^4$
\kmsec).  The forward shock produces a very hot shell ($\sim$10$^9$
K), while the reverse shock produces a denser, cooler shell
($\sim$10$^7$ K) with much higher emission measure, from which the
observable X-ray emission arises (Chevalier 1982).  This picture is
modified if the stellar wind is clumpy; then the forward shock front
in the clumps can yield cooler, stronger emission than the shock front
in the diffuse wind (Chugai 1993).

The exact time of the explosion of \snn\ in the host galaxy \mcg\ is
not known, but it may have occurred as much as ten months before its
optical discovery in May 1995 (Benetti, Bouchet \& Schwarz 1995),
making the supernova $\sim$2 years old at the epoch of first detection
(Lewin, Zimmermann, \& Aschenbach 1996).  \snn\ was only the fourth
supernova to be detected in X-rays so soon after its explosion.
Regular observations of the X-ray temperature, luminosity, and
line-of-sight hydrogen column density will help determine the
structure of the SN ejecta, the structure of the circumstellar medium
established by the wind of the progenitor, and the details of its
evolution prior to the explosion.  These constraints can lead to an
estimate of the progenitor stellar type and the type and length of its
(pre-explosion) evolutionary stages.

\section{Observations}
\label{sec:obs}
We first observed \snn\ in July 1996 with \rosatz\ when the optical
(Benetti \etal\ 1995; Garnavich \& Challis 1995) and radio (Van Dyk
\etal\ 1996) properties of this supernova made it a promising
candidate for strong X-ray emission.  X-ray detection of \snn\ with
this pointing (Lewin \etal\ 1996) prompted our longer follow-up
observation that August.  In addition, we were awarded time for
another \rosatz\ observation in August 1997, and applied
simultaneously for \ascaz\ spectral observations.  Execution of our
\ascaz\ observation in January 1998 completed the program for this
paper.

The log of our X-ray observations of \snn\ is presented in
Table~\ref{tbl:obs}.  
\begin{table}
\caption{Log of \snn\ X-ray observations}
\begin{tabular}{r@{~}r@{~}lcrrr}
  ~ &   ~ &    ~ &              & Exp.  &      ~            & \ccol{Rate}  \\
 \multc{3}{Date} & Instrument   & (sec) & \ccol{Counts}     & (ksec$^{-1}$)\\
\hline
 23 & Jul & 1996 & \rosatz\ HRI &  1311 &    9.4$\pm\,$~0.6 &  7.4$\pm$3.5 \\
 12 & Aug & 1996 & \ccol{''}    & 16991 &  163.6$\pm\,$~1.9 &  9.9$\pm$0.9 \\
 17 & Aug & 1997 & \ccol{''}    & 18797 &  126.4$\pm\,$~1.9 &  6.9$\pm$0.7 \\
 20 & Jan & 1998 & \ascaz\ SIS  & 91129 & 1960.4$\pm$29.6   & 29.7$\pm$1.1 \\
  ~ &   ~ &    ~ & \ascaz\ GIS  & 95944 & 1300.7$\pm$15.9   & 18.3$\pm$0.8 \\
\hline
\end{tabular}
\medskip

Exposures are given in seconds, and count rates are in counts per
kilosecond; \ascaz\ `SIS' and `GIS' figures are for the SIS0+SIS1 and
GIS2+GIS3 detector combinations, respectively (with the rates averaged
between detectors).  The quoted uncertainty in the number of counts
detected is the uncertainty in the background subtraction only and
does not take account of Poisson statistics.  Count rates are
corrected for the effects of the point spread function, and count rate
uncertainties include the Poisson contribution.
\label{tbl:obs}
\end{table}
The standard photon screening criteria were adopted for all data sets
(David \etal\ 1995; ASCA Data Reduction Guide\footnote{http://heasarc.gsfc.nasa.gov/docs/asca/abc/abc.html}).  Source and background count
rates were determined as follows.  

\rosatz\ source counts were taken from a circle of radius 17.5~arcsec
(35 pixels) centred on the SN position, and background counts were
taken from an annulus of inner radius 20~arcsec and outer radius
50~arcsec.  Source count rates (but not raw count numbers) were
corrected for incompleteness due to the wings of the instrument point
spread function (PSF), a 3\% effect.

\begin{figure}
\centering
~\psfig{file=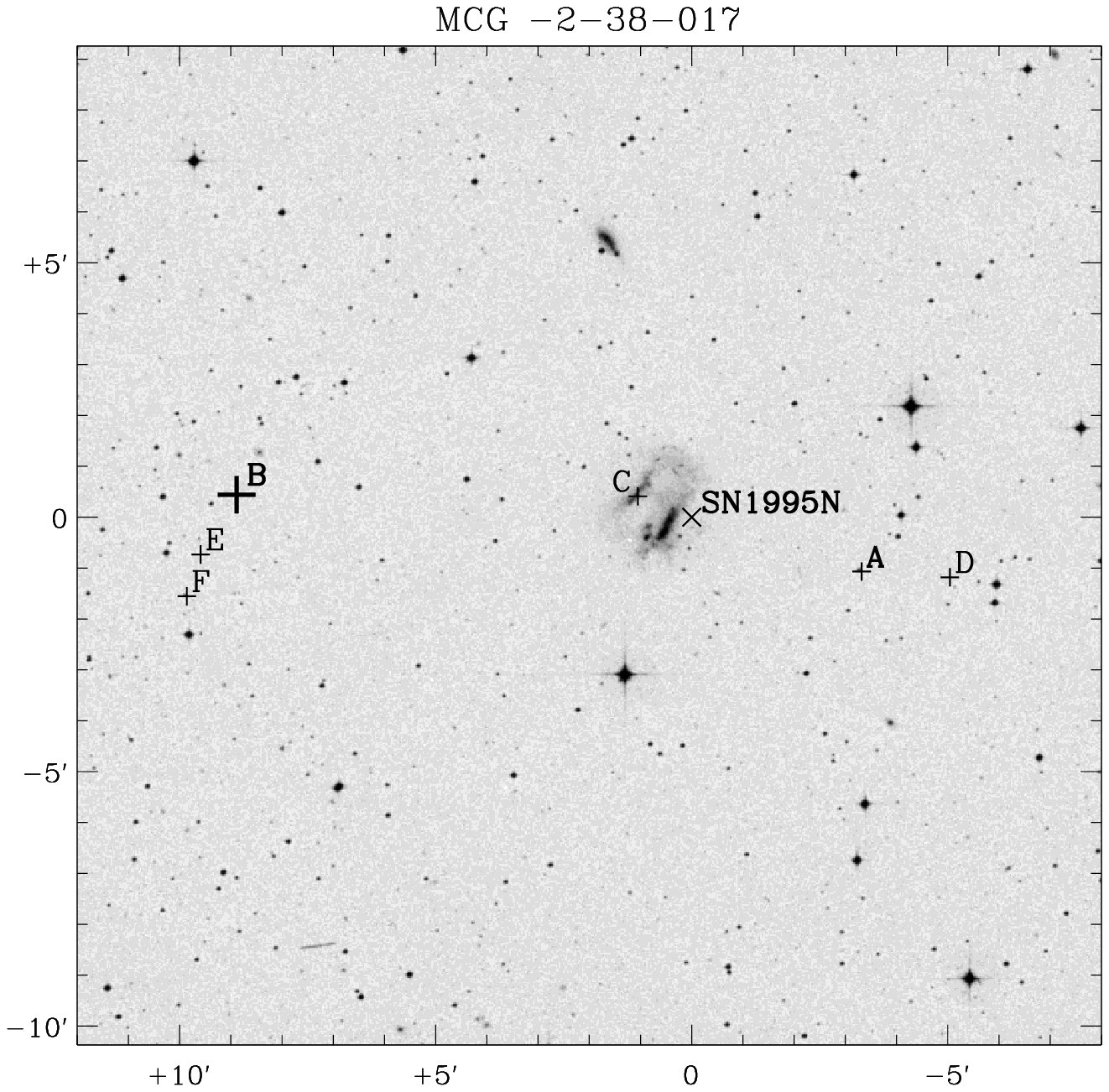,angle=90,width=8cm}~
\caption{The \snn\ field in the optical ($J$~band, from the Digitized
Sky Survey), showing the host galaxy \mcg.  Coordinates are in
arcminutes relative to the position of \snn, J2000 R.A. 14\rah 49\ram
28\ras 313, Dec $-$10\dcd 10\dcm 13\dcs 919 (indicated; Van Dyk \etal\
1996).  Positions of the nearby (\asca) sources `A' and `B', and the
additional (\rosat) sources `C'--`F' are also shown.  The \ascaz\ GIS
source `B' is not detected in our \rosatz\ pointings, and therefore
its positional uncertainty is substantial ($\sim$0.5 arcmin). }
\label{fig:dss}
\end{figure}

Two nearby X-ray sources complicated the process of choosing source
and background extraction regions for the \ascaz\ images.  In
Figure~\ref{fig:dss} we show the positions of \snn\ and these X-ray
sources, `A' and `B', overlaid on an optical image of the area
from the Digitized Sky Survey.  Also shown are four additional, weaker
sources (`C'--`F'), visible in the summed HRI image, that are
ignored in the \ascaz\ analysis.

\ascaz\ SIS source counts were taken from a circle of radius 4~arcmin
(38 pixels) centred on the SN position, excluding a smaller circle
(radius 15 pixels) around source A.  Background counts were taken from
the area of the CCD that remained after exclusion of the SN (38 pixel
radius) and source A (24 pixel radius), and the SN count rate was
corrected for the fraction of the PSF excluded (with the source area
used expected to include 73\% of the source counts).

\ascaz\ GIS source counts were taken from a circle of radius
4.9~arcmin (20 pixels) centred on the SN position, excluding a smaller
circle (radius 7 pixels) chosen on the basis of the SIS image to
exclude counts from source A.  Background counts were taken from a
larger circle of radius 11.8~arcmin (48 pixels) centred on the SN,
with circular areas removed to account for the SN (24 pixel radius),
source A (18 pixel radius) and source B (20 pixel radius).  The source
area used here is expected to include 74\% of the source counts.

\section{Analysis}
\label{sec:analysis}

\subsection{Spectrum}
\label{sub:anal-spectrum}
Our X-ray spectral analysis made use only of the \ascaz\ data, since
the HRI does not have a spectral response sufficient for our purposes.
After extracting source and background counts (see Sec.~\ref{sec:obs}),
we added the data from the two instrument pairs (SIS0+SIS1 and
GIS2+GIS3) and produced averaged instrument response matrices for our
analysis.  We then `grouped' spectral channels to achieve a minimum of
$\sim$20 counts per channel above background.

X-ray spectral fitting was performed within the XSPEC environment
(Arnaud 1996).  We used the default modified Levenberg-Marquardt
algorithm (derived from CURFIT, Bevington 1969) to minimize a $\chi^2$
statistic that was calculated using the weighting scheme of Gehrels
(1986); the statistical error on $N$ counts was taken to be $1 +
\sqrt{N + 0.75}$.

SIS (0.5--7.0~keV) and GIS (1.0--7.0~keV) spectra were fit
simultaneously, with only the relative normalization of the model
allowed to vary between the two datasets.  This allowed for
differential errors in our treatment of background subtraction and PSF
effects between the two sets of instruments.

The quality of the spectrum (2000 SIS photons and 1300 GIS photons)
does not make the examination of complex spectral models necessary or
meaningful.  On the contrary, as shown in Table~2,
power-law and thermal bremsstrahlung spectra both provide acceptable
fits to the data.
\begin{table*}
\label{tbl:xspec}
\begin{minipage}{103mm}
\caption{\snn\ X-ray Spectral Fits}
\begin{tabular}{lcccrr}
    ~   &        ~       & \nhcol             & ~ & \multc{2}{Flux (0.5--7 keV)} \\
  Model & $\chi_{\nu}^2$ &($10^{21}$ cm$^{-2})$ & Parameter & 
  \ccol{SIS} & \ccol{GIS} \\
\hline
  Power Law & 0.956 & 1.8$\pm$0.5 & $\alpha = 1.7\pm 0.1$ & 
             11.1$\pm$1.2 & 12.8$\pm$1.4 \\
  Bremss.   & 0.999 & 1.1$\pm$0.4 & $kT = 9.1^{+2.7}_{-1.8}$ & 
             11.1$\pm$0.6 & 12.7$\pm$0.7 \\
\hline
\end{tabular}
\medskip

Absorbed 0.5--7~keV flux is given in units of 10$^{-13}$ \ergcms, and
is corrected for incompleteness due to the instrument point spread
functions.  The power-law index $\alpha$ is the photon spectral index.
Reduced chi-squared values are calculated using a Gehrels weighting
(Gehrels 1986).  Note that the Galactic neutral hydrogen column in the
direction of \snn\ is $7.8 \times 10^{20}$ \pcmsq\ (Dickey \& Lockman
1990).
\end{minipage}
\end{table*}
MEKAL (Mewe, Gronenschild \& van den Oord 1985) and Raymond-Smith
plasma models also provide good fits to the data, but with the
abundance parameter of these models essentially unconstrained in the
fits, it is not clear what further conclusions the models allow us to
draw.  Plasma temperatures in the fits are $kT =
6.9^{+1.4}_{-0.9}$~keV (MEKAL) and $kT = 8.8^{+2.9}_{-1.6}$~keV
(Raymond-Smith), with hydrogen column densities $\nhcol \sim
10^{21}$~\pcmsq.  

By contrast, a simple blackbody model (with absorption) was unable to
fit the data and can be excluded -- by comparison with the power law
or bremsstrahlung models -- with 99.9\% confidence.

\begin{figure}
\centering
~\psfig{file=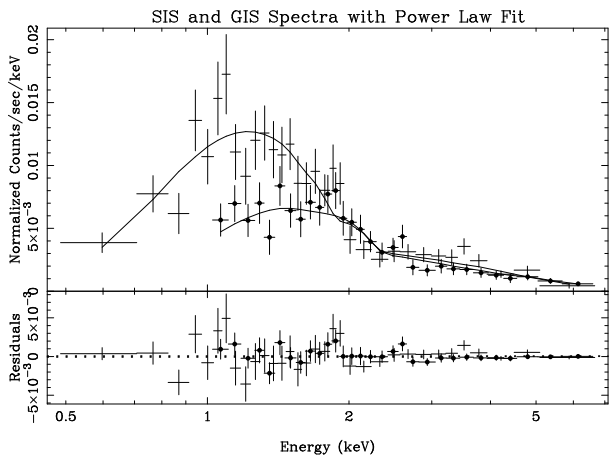,width=8cm}~\\
\caption{The X-ray spectrum of \snn, with power-law model fit.
 The fitted power-law exponent (photon index) is $\alpha = 1.7\pm 0.1$
 with an absorbing column $\nhcol = 1.8\pm 0.5 \times 10^{21}$~\pcmsq;
 the spectrum may also be fit with a thermal bremsstrahlung model with
 $kT = 9.1^{+2.7}_{-1.8}$~keV (see Table~2).  The
 residuals show a `bump' near 1.8~keV, an appropriate energy for
 fluorescent Si emission, but the fit is not significantly improved by
 addition of an emission line at that energy.}
\label{fig:xspec-pl}
\end{figure}

Figure~\ref{fig:xspec-pl} shows the observed X-ray spectrum with the
power law fit superposed (the bremsstrahlung model fit does not differ
visibly from this).
Although there is a hint in the fit residuals of an additional feature
near 1.8~keV, addition of an absorption edge or spectrally unresolved
emission line at this energy cannot be justified by the statistics
(F-statistic confidence levels of 54\% and 62\%, respectively).
Nevertheless, the energy is appropriate for fluorescent silicon
emission (Si~$K\alpha$ and $K\beta$ neutral line energies are 1.74~keV
and 1.84~keV, respectively) and it is possible that the feature is
real; if observations of other X-ray SNe with \chandra\ and \xmm\ are
able to detect a feature at this energy then the interpretation of our
data will have to be revisited.  

We note that the fitted hydrogen column densities under the two models
are consistent with each other and with the Galactic column towards
\snn, which is $7.8 \times 10^{20}$~\pcmsq\ (Dickey \& Lockman 1990).

We examined the data near 6.7~keV for evidence of iron line emission
and find none: our 90\%-confidence upper limit for the equivalent
width of a 6.7~keV iron line is 1.1 keV.  Upper limits to the
equivalent widths of lines centred at energies from 6.2 to 7.2 keV
are all similarly $\sim$1~keV.

\subsection{Light Curve}
\label{sub:anal-ltcurve}
With an X-ray spectrum in hand we can proceed to calculate the
incident flux and total X-ray luminosity of \snn\ at the several
epochs of our observations.  These calculations are made under the
assumption that (1) the spectrum of the source (apart from its overall
normalization) was roughly constant over the timespan of the
observations; and (2) that it is sufficiently well described by our
models to allow extrapolations outside the energy range of our
\rosatz\ and \ascaz\ data. Note that in 1996 and 1997 we have only
\rosatz\ data (no spectral information), and in 1998 we have only
\ascaz\ data (no sensitivity below 0.5~keV) -- all other quoted fluxes
and luminosities are based on model extrapolations.

Using the spectral fits as derived for the {\it ASCA} data of 1998
(see Sect.\ 3.1) and the PIMMS v3.0 software
package\footnote{http://heasarc.gsfc.nasa.gov/docs/software/tools/pimms.html},
we calculate the unabsorbed fluxes corresponding to the count rates
observed in our \rosatz\ data, adding together the July and
August~1996 pointings for better statistics; HRI count rates for the
two observations are consistent with no change over the intervening
20~days (Table~\ref{tbl:obs}).  Independently, we extrapolate our 1998
\ascaz\ fluxes to the 0.1--2.4~keV range of the HRI for comparison
purposes; the results are reported in Table~3.
\begin{table}
\label{tbl:flux}
\caption{X-ray flux history of \snn}
\begin{tabular}{cccccl}
  ~   &           ~   &\multc{2}{Unabsorbed Flux} & ~      & Stat.\\
 Date & Instr.& 0.1--2.4 keV & 0.5--7.0 keV & \Lx  & Unc.\\
\hline
 Jul--Aug~1996 & HRI & 6.3--8.9 & \dgsp 9.9--11.0 & 12 & \dgsp 9 \\
   17~Aug~1997 & HRI & 4.5--6.3 &  7.0--7.8        & \dgsp 8 & 10 \\
   20~Jan~1998 & SIS &\dgsp 7.5--10.5 & 12.2--13.2 & 14 & \dgsp 5 \\
             ~ & GIS &\dgsp 8.6--11.5 & 14.0--15.2 & 15 & \dgsp 5 \\
\hline
\end{tabular}
\medskip

Unabsorbed fluxes are in units of 10$^{-13}$ \ergcms, within the
energy ranges indicated.  Luminosities \Lx\ (0.1--10~keV; assumed
isotropic) are in units of $10^{40}$ \ergsec, and are calculated using
the Galactic hydrogen column $\nhcol = 7.8\times 10^{20}$~\pcmsq\ and
a distance of 28~Mpc.  Flux ranges indicate the variation in
photon-to-flux conversion for our two best-fit spectral models,
thermal bremsstrahlung (left) and power-law (right), due mainly to the
difference in the fitted hydrogen column (see
Sec.~\ref{sub:anal-spectrum}).  The additional percentage error
introduced by the statistical uncertainty in the number of photons
detected for each observation is shown.  Since the hydrogen column
measurement has additional statistical and systematic uncertainties,
these ranges are likely to be overoptimistic (see text for details).
\end{table}
Since our power-law and thermal bremsstrahlung spectral models have
different best-fit column densities, and this uncertainty dominates
the Poisson uncertainty in our unabsorbed flux determinations, we
report the flux conversions for the two models separately (in terms of
a low and a high value) in Table~3.  It should be noted, however, that
in all likelihood the reported ranges still underestimate the actual
uncertainty in unabsorbed source flux, since (1) the value of the
best-fit hydrogen column for both models is relatively uncertain
(Table~2); and (2) our spectral fits are subject to additional,
unquantified, systematic uncertainties due to imperfect background
subtraction.  For example, if we reduce the hydrogen column to the
Galactic value of $7.8\times 10^{20}$~\pcmsq\ -- within the 1$\sigma$
range for our bremsstrahlung fit, and the 2$\sigma$ range for our
power-law fit -- then the fluxes calculated from our HRI observations
are reduced by a further 10\% relative to the lower end of the ranges
given in Table~3.

Accepting these caveats, we may draw the following conclusions.
First, the \rosatz\ observations indicate (at $>$4-sigma confidence) a
30\% decrease in absorbed 0.1--2.4~keV X-ray flux between August~1996
and August~1997 (Table~1).  Second, comparison of the August~1997
\rosatz\ observation with the January~1998 \ascaz\ observation
(Table~3) indicates that the X-ray luminosity of \snn\ may have
increased by a factor of $\approx$2 over this time.  Alternatively,
the spectrum may have changed significantly (contrary to our
Assumption~1 above); however, to make the August~1997 observation
consistent with $\Lx = 1.5\times 10^{41}$~\ergsec\ requires either
absorption of $\nhcol \sim 4\times 10^{21}$~\pcmsq\ or an unlikely
(temporary) hardening of the spectrum to $kT > 50$~keV or $\alpha <
1.02$.  Thus, the conclusion that the SN dimmed from August~1996 to
August~1997 and then brightened from August~1997 to January~1998 seems
the most probable explanation for our data.

\section{Discussion}
\label{sec:discuss}

The optical spectra of \snn\ show that it belongs to the Type~IIn
(narrow line) category (Benetti \etal\ 1995).  A number of other
Type~IIn supernovae have been detected as strong X-ray sources and
their properties are summarized in Table~4.  These supernovae are the
most luminous X-ray supernovae and are inferred to be interacting with
a dense circumstellar medium.  They are also luminous radio
supernovae.  The X-ray luminosity of \snn\ places it in this highly
luminous group.

\begin{table*}
\label{tbl:IIn}
\begin{minipage}{91mm}
\caption{Type IIn X-ray Supernovae}
\begin{tabular}{cccccl}
       ~    &   Distance & $\log \Lx$ &   $kT$   & $\log \nhcol$ & ~  \\
  Supernova & (Mpc)    &(ergs s$^{-1}$) & (keV) & (cm$^{-2}$) & Refs. \\
\hline
 SN 1978K & 4.5 & 40.3 & 3        & $\approx$20   & (1),(2) \\
 SN 1986J & 10  & 40.3 & 5.0--7.5 & 21.7 & (3)  \\
 SN 1988Z & 98  & 41.0 &          &      & (4)  \\
 SN 1995N & 28  & 41.2 & 9        & 20.9 & (5)  \\
\hline
\end{tabular}
\medskip

\Lx\ is the peak observed X-ray luminosity (0.1--10~keV).  References
are: (1) Petre \etal\ 1994; (2) Schlegel \etal\ 1996; (3) Houck \etal\
1998; (4) Fabian \& Terlevich 1996; (5) this paper.
\medskip
\end{minipage}
\end{table*}

The interpretation of the X-ray emission from these sources is not
clear.  Chevalier \& Fransson (1994) presented models in which smooth
supernova ejecta interact with a smooth circumstellar wind.  In this
picture, the reverse shock emission is characterized by $kT\approx
1$~keV and the forward shock emission by $kT\approx 100$~keV.
However, the radio emission (see Weiler \etal\ 1990 for SN~1986J) and
narrow optical line emission give evidence that the circumstellar
medium is clumpy and shock waves are being driven into the clumps (see
Chugai \etal\ 1995 on the case of SN~1978K).  Supernovae like SN~1978K
and SN~1988Z show very narrow line emission, which can be attributed
to dense circumstellar clumps that are photo-ionized by the X-ray
emission, and line components with a width of $\sim 2,000$~\kmsec\
that can be from cooling shocks moving into the clumps.  \snn\ also
shows evidence for these line components (A.V.\ Filippenko,
priv. comm.).  Shock waves in clumps can give cooler X-ray emission
than the forward shock front in the diffuse circumstellar medium.
Chugai (1993) found that this picture may be preferred for the X-ray
emission from SN~1986J.  A prediction of this model is that the
emitting region should have a lower characteristic velocity than in
the case of the reverse shock wave.

Our series of flux measurements show that the X-ray luminosity of
\snn\ has probably dimmed by 30\% and then brightened by as much as a
factor of two over the period of observation, which corresponds to an
age range of 2.0 to 3.5 years for a proposed explosion date in
July~1994 (Benetti \etal\ 1995).  The X-ray light curve of SN~1978K
shows approximately constant luminosity over the age range of 12 to 16
years (Schlegel, Petre, \& Colbert 1996).  For SN~1986J, a decline in
luminosity has been observed over the age range 8.6 to 13 years (Houck
\etal\ 1998); the evolution is consistent with a $t^{-2}$ time
dependence.  In the models of Chevalier \& Fransson (1994), a decline
in flux is expected if the evolution is non-radiative, whereas a
radiative reverse shock front leads to a roughly constant luminosity.
In the clump model (Chugai 1993), the evolution depends on the
variation of clump properties with radius.  The light curve that we
observe may be due to such variation.

Characterization of the radio observations of \snn\ is not yet
complete.  However, five 8.4~GHz radio observations made between
July~1995 and June~1998 are all consistent with a constant flux of
3.8~mJy (Van Dyk \etal\ 2000), indicating only mild evolution of the
radio properties of the shocked region over this interval.  

In Table~\ref{tbl:IIn}, we give the estimated temperature
characterizing the X-ray emission, although it is only in the case of
SN~1986J that an X-ray line has been detected (Houck \etal\ 1998), and
hence, the presence of thermal emission confirmed.  Possible
non-thermal sources of X-ray emission are inverse Compton and
synchrotron emission, but these are typically less efficient than
thermal emission (Chevalier 1982).  The temperature that we find is
higher than expected for the reverse shock wave, but somewhat lower
than that expected for the forward shock wave.  If the emission is
from shocked clumps, the temperature depends on the density of the
clumps, so that intermediate temperatures are possible.  The lack of
detectable line emission does not allow us to directly test for the
presence of clumps by examining the widths of lines.

The gas column density that we measure to the source is compatible
(within our roughly 50\% uncertainty) with the expected column density
in our Galaxy.  Table \ref{tbl:IIn} shows that a higher column density
is present to SN~1986J, but that can be attributed to absorption
within the parent galaxy (Houck \etal\ 1998).  These supernovae do not
show evidence for absorption by the dense shell that might be formed
by a cooling reverse shock wave.  Since the cool shell formed by a
reverse shock will probably not be ionized, two years or more after
the SN explosion, this low absorption may indicate inhomogeneities in
the shell that allow the X-ray flux to emerge without absorption.
Alternatively, low absorption may be seen as supporting the clump
model (Chugai 1993), since the X-ray emission from forward shocks
penetrating high-density clumps will not be significantly absorbed.

\section{Conclusions}
\label{sec:conclusions}
Our observations show that \snn\ belongs to the class of X-ray
luminous Type~IIn supernovae.  The variations in X-ray flux over a
1.5~yr period suggest that the emission mechanism is more complicated
than simple radiative emission from a reverse shock wave generated by
interaction with a dense circumstellar medium, and that
inhomogeneities in the circumstellar medium play a role.  The X-ray
emission is expected to power the optical/ultraviolet emission from
the supernova; thus, optical spectroscopy will provide information on
the gas motions so that a model for the supernova and the
circumstellar structure can be determined.

\subsection{Acknowledgments}
WHGL and RC gratefully acknowledge support from the National
Aeronautics and Space Administration. KWW thanks the office of Naval
Research for the 6.1 support of this research.  This research has made
use of the Digitized Sky Survey, produced at the Space Telescope
Science Institute under U.S. Government grant NAG W-2166.

\label{lastpage}


\begin{thebibliography}{}

 \bibitem{Arnaud-96}Arnaud K., 1996, Astronomical Data Analysis
  Software and Systems V, Eds.\ Jacoby G., Branes J., p.\ 17, ASP
  Conf.\ Series Vol.\ 101

 \bibitem{Benetti-et-95}Benetti S., Bouchet P., Schwarz H., 1995, IAUC 6170 

 \bibitem{Bevington-69}Bevington R., 1969, Data Reduction and Error
  Analysis for the Physical Sciences, McGraw-Hill, New York

 \bibitem{Chevalier82}Chevalier R.A., 1982, ApJ, 259, 302

 \bibitem{Chevalier-Fransson-94}Chevalier R.A., Fransson C., 1994, ApJ,
  420, 268

 \bibitem{Chugai-93}Chugai N.N., 1993, ApJ, 414, L101
 
 \bibitem{Chugai-et-95}Chugai N.N., Danziger, I. J., Della Valle, M.,
  1995, MNRAS, 276, 530

 \bibitem{David-et-95}David L. P., Harnden Jr.\ F. R., Kearns K. E.,
  Zombeck M. V., 1995, The ROSAT High Resolution Imager.  U.S. ROSAT
  Science Data Center/SAO, Cambridge, MA

 \bibitem{Dickey-Lockman-90}Dickey J.M., Lockman F.J., 1990, ARAA, 28,
  215 

 \bibitem{Fabian-Terlevich-96}Fabian A.C., Terlevich R., 1996, MNRAS,
  280, L5 

 \bibitem{Fox-Lewin-99}Fox D.W., Lewin W.H.G., 1999, IAUC 7318

 \bibitem{Garnavich-Challis-95}Garnavich P., Challis P., 1995, IAUC 6174 

 \bibitem{Gehrels-86}Gehrels N., 1986, ApJ, 303, 336

 \bibitem{Houck-et-97}Houck J.C., Bregman J.N., Chevalier R.A.,
  Tomisaka K., 1998, ApJ, 493, 431
  
 \bibitem{Immler-et-98a}Immler, S., Pietsch, W.,  Aschenbach, B. 1998a,
  A\&A, 331, 601

 \bibitem{Immler-et-98b}Immler, S., Pietsch, W.,  Aschenbach, B. 1998b,
  A\&A, 336, L1

 \bibitem{Lewin-et-96}Lewin W., Zimmermann H.U., Aschenbach B., 1996, IAUC 6445
  
 \bibitem{Matzner-et-99}Matzner, C.D., McKee C.F., 1999, ApJ, 510, 379

 \bibitem{Mewe-et-85}Mewe R., Gronenschild E.H.B.M., van den Oord
  G.H.J., 1985, A\&AS, 62, 197

 \bibitem{Petre-et-94}Petre R., Okada, K., Mihara, T. Makishima, K,
   Colbert, E. J. M., 1994, PASJ, 46, L115

 \bibitem{Pollas-95}Pollas C., 1995, IAUC 6170

 \bibitem{Schlegel-95}Schlegel, E. M. 1995, Rep. Prog. Phys. 58, 1375
 
 \bibitem{Schlegel-et-96}Schlegel, E. M., Petre, R., Colbert, E. J. M.
   1996, ApJ, 456, 187

 \bibitem{VanDyk-et-96}Van Dyk S.D., Sramek R.A., Weiler K.W., Montes
  M.J., Panagia N., 1996, IAUC 6386

 \bibitem{VanDyk-et-00}Van Dyk S.D., \etal, 2000, in preparation

 \bibitem{Weiler-et-90}Weiler K.W., Panagia N., Sramek R.A., 1990, ApJ,
  364, 611

\end{thebibliography}
\end{document}